\documentclass[%
 reprint,
superscriptaddress,
 amsmath,amssymb,
 aps,
]{revtex4-2}

\usepackage{aas_macros}
\usepackage{pifont}
\usepackage{graphicx}
\usepackage{dcolumn}
\usepackage{bm}
\usepackage{hyperref}

\usepackage{subfigure}%
\usepackage{float}%
\hypersetup{
     colorlinks   = true,
     citecolor    = blue
}

\hypersetup{colorlinks=true, citecolor=blue, linkcolor = magenta}

\begin{document}

\preprint{APS/123-QED}

\title{Shadow and gravitational lensing produced by the nonlinear accretion of a scalar field onto a black hole}

\author{J. C. Acevedo-Muñoz}
\email{jennyfer2238304@correo.uis.edu.co}
\affiliation{Grupo de Investigaci\'on en Relatividad y Gravitaci\'on, Escuela de F\'isica, Universidad Industrial de Santander A. A. 678, Bucaramanga 680002, Colombia}
 
\author{F.~D.~Lora-Clavijo} 
\email{fadulora@uis.edu.co}
\affiliation{Grupo de Investigaci\'on en Relatividad y Gravitaci\'on, Escuela de F\'isica, Universidad Industrial de Santander A. A. 678, Bucaramanga 680002, Colombia}

\author{A. Cruz-Osorio}
\email{aosorio@astro.unam.mx}
\affiliation{Instituto de Astronom\'{\i}a, Universidad Nacional Aut\'onoma de M\'exico,\\ AP 70-264, 04510 Ciudad de M\'exico, M\'exico}


\date{\today}

\begin{abstract}
The hypothesis that classical scalar fields could constitute dark matter on galactic and cosmic scales has garnered significant interest. In scenarios where supermassive black holes (SMBHs) form through the accretion of matter onto black hole seeds, a critical question arises: what role does dark matter play in this process? We conduct a numerical investigation into the nonlinear dynamical evolution of black hole shadows and gravitational lensing effects resulting from the accretion of an ultralight, real scalar field onto a non-rotating black hole. The scalar field is minimally coupled to Einstein’s gravity, and our simulations focus on wave packets, parameterized by their wave number and width, as they interact with and are accreted by a dynamic black hole. Our results demonstrate significant growth in the apparent horizon, photon ring, and Einstein ring sizes compared to those of a Schwarzschild black hole. These findings suggest that the observed photon ring and black hole shadow in  Sgr\,A*, and M87* may be influenced by the gravitational interaction between the black hole and ultralight scalar field dark matter. 
\end{abstract}

\maketitle


\section{\label{sec:intro} Introduction}
Nowadays, dark matter is one of the most important open problems in astrophysics, and discovering its nature is possibly one of the greatest challenges of modern physics. Currently, the most widely accepted model of dark matter in the scientific community is the ``Cold Dark Matter" (CDM) \cite{1996ApJ...462..563N}, which successfully explains phenomena such as observations of the cosmic microwave background and the large-scale distribution of structure \cite{2003MNRAS.346L..26E, 2004Natur.427...45B}. However, discrepancies have emerged at the galactic scale, such as the core-cusp problem in the density profile of galactic halos \cite{1994Natur.370..629M, 2015AJ....149..180O}. These challenges have prompted the exploration of alternative dark matter models.

One promising alternative is the Scalar Field Dark Matter (SFDM) model, which posits that dark matter could be composed of ultralight scalar fields. This model has been proposed to address issues at both galactic \cite{2000CQGra..17L...9S, 2000PhRvD..62f1301M} and cosmological scales \cite{2000PhRvD..62j3517S,2000CQGra..17.1707M,matos2000quintessence}. The main idea behind SFDM is that the properties of dark matter can be represented by a scalar field with an extremely low mass of approximately $10^{-22} eV$  \cite{2019FrASS...6...47U}. The SFDM model offers a compelling alternative to CDM, providing a potential resolution to the core-cusp problem and other small-scale challenges, while opening new avenues for understanding the fundamental nature of dark matter.

The idea of scalar fields as a candidate for dark matter was first proposed in 1983, when researchers demonstrated that both bosonic and fermionic particles could potentially account for the observed galactic rotation curves, establishing one of the earliest theoretical links between scalar fields and dark matter \cite{1983PhLB..122..221B}. This model gained further traction when galactic rotation curves were successfully fitted using a Bose gas model, suggesting the existence of an ultralight scalar field dark matter particle \cite{PhysRevD.50.3650, PhysRevD.50.3655}. Subsequent advances included numerical simulations of galaxy formation using scalar fields, where the Schrödinger-Poisson system was employed to model the structure of galaxies computationally \cite{1993ApJ...416L..71W}.

In 1998, a significant breakthrough emerged with the proposal that dark matter could be described as a scalar field. This work demonstrated that the scalar field hypothesis could effectively explain the observed rotation patterns of stars and gas in spiral galaxies \cite{2000CQGra..17L...9S}. Further studies utilized the Thomas-Fermi approximation to analyze rotation curves, revealing distinct density profiles for the SFDM model. For real scalar fields, the density profile follows a sine function concerning the distance ratio \cite{1996PhRvD..53.2236L, 2011JCAP...05..022H}, while for complex scalar fields, it exhibits a squared sine relationship \cite{2013PhRvD..88h3008R}. Additionally, alternative density profiles involving power-law relationships have been proposed through numerical simulations \cite{2014NatPh..10..496S}. Beyond density profiles, the SFDM model provides a natural explanation for the baryonic Tully-Fisher relation, further strengthening its viability as a dark matter candidate \cite{2019arXiv190100305L, bray2014wave}.

On a cosmic scale, the SFDM model has emerged as a compelling candidate for explaining dark matter in large-scale structures, particularly when its theoretical predictions are rigorously compared with key cosmological observations. Notably, the model demonstrates consistency with data from the Cosmic Microwave Background (CMB) and the Matter Power Spectrum (MPS), aligning well with the observed distribution and evolution of matter in the universe \cite{2001PhRvD..63f3506M, 2011MNRAS.413.3095H}.

A distinctive feature of the SFDM model is its proposal that dark matter could consist of ultralight scalar particles in a Bose-Einstein condensate state. Due to their extremely low mass, these particles behave collectively as a coherent wave, introducing quantum pressure effects that suppress the formation of small-scale structures. This property provides natural resolutions to persistent issues in the CDM model, such as the core-cusp problem and the missing satellite problem \cite{2012MNRAS.422..282R, 2011JCAP...05..022H, 2015MNRAS.450..209B}.

The SFDM model is particularly appealing because it not only reproduces the successes of the CDM model at large scales but also offers a plausible explanation for the observed structure formation in the universe \cite{2000CQGra..17L..75M, 2001PhRvD..63f3506M, 1990PhRvL..64.1084P, PhysRevD.50.3650}. These characteristics make the SFDM model a robust and versatile framework for understanding the nature of dark matter across both galactic and cosmological scales. For further details on the SFDM model, including its theoretical framework and cosmological applications, see \cite{2024FrASS..1147518M}.

The interplay between scalar fields and black holes is fundamental for comprehending the physical phenomena related to the formation and evolution of supermassive and intermediate black holes \cite{2014MNRAS.443.2242L, 10.1046/j.1365-8711.2003.06232.x}. While the prevailing notion suggests that baryonic matter constitutes the primary source of mass accretion onto black holes, it is important to consider scenarios where dark matter might play a significant role in the accretion process \cite{1977ApJ...211..244L, 2002NewA....7..385Z, 2008PhRvD..77f4023P, 2010MNRAS.404L...6H}. Scalar field accretion onto black holes has already been explored in different works. Within the context of ultra-light scalar fields proposed as constituents of dark matter halos, it has been demonstrated that the accretion rate remains low over the typical lifespan of a halo. Consequently, central supermassive black holes could potentially coexist with these scalar-field halos \cite{2002PhRvD..66h3005U}. The study of scalar-field accretion has also been extended to the spacetime backgrounds of boson stars \cite{2010PhRvD..82b3005L} and black holes \cite{2011JCAP...06..029C,Cruz2010a}, employing the ``Scri''  ($\mathcal{I}$)-fixing conformal compactification method. In \cite{2011PhRvD..83d3525B,2012PhRvL.109h1102B}, it was demonstrated that a massive scalar field surrounding a Schwarzschild black hole can persist for extended periods by forming quasi-bound states. Furthermore, within the framework of the nonlinear evolution of massless scalar fields, accounting for spacetime evolution, in \cite{PhysRevD.85.024036} revealed that scalar fields can persist outside black holes, potentially with lifetimes consistent with cosmological timescales. Subsequent studies have corroborated the existence of such long-lived states \cite{2013PhRvD..87d3513W,2015PhRvD..91d3005S,2017PhRvD..96b4049B,2023PhRvD.107d4070A}. Equilibrium states between ultralight scalar fields and black holes in general relativity were found in \cite{PhysRevLett.112.221101,2015CQGra..32n4001H}. 
 On the other hand, the effects of ultralight bosons on the dynamics of a SMBH shadow induced by superradiance have been estimated using semi-analytical models, introducing two potential observables: shadow drift and photon ring autocorrelation parameter \cite{Chen2022}.

Observations from the Event Horizon Telescope (EHT) Collaboration revealed the first image of a supermassive black hole at the core of the elliptical galaxy M87 \cite{EHT_M87_PaperI,EHT_M87_2018_2024}, followed by the image of the supermassive black hole at the galactic center, Sgr\,A* \cite{EHT_SgrA_PaperI}. These images provided the first confirmation of a black hole's shadow and photon ring, consistent with theoretical predictions from general relativity.
The observable features of the black hole, including the shadow and photon ring size, now serve as crucial quantities for testing theories of gravity, plasma physics, particle acceleration, and the nature of dark matter. Early models of the dark matter halo surrounding the M87 galaxy have shown that the black hole shadow size remains very close to the vacuum Kerr solution \cite{Jusufi2019,Liu2024}. However, in these models, dark matter is incorporated by modifying the Kerr solution, without including feedback mechanisms or evolution equations for dark matter. 

This paper presents a numerical analysis of the nonlinear dynamical evolution of the shadow and gravitational lensing effects resulting from the accretion of a scalar field onto a black hole. In Section \ref{sec:ESFSE}, we provide a concise overview of the governing equations for the evolution of the scalar field and spacetime geometry, along with a detailed discussion of the numerical methods employed to solve these equations. The black hole shadow and gravitational lensing results are presented and analyzed in Section \ref{sec:results}. Finally, in Section \ref{sec:discussion}, we summarize our findings, discuss their implications, and outline potential avenues for future research. 

Throughout this work, we adopt the metric signature $(-,+,+,+)$ and use geometrized units, where the gravitational constant $G$ and the speed of light $c$ are set to unity.

\section{Einstein-scalar field system of equations} \label{sec:ESFSE}

We numerically solve Einstein's equations by splitting spacetime using the 3+1 formalism of general relativity and the ADM equations to evolve the components of the spatial metric, $\gamma_{ij}$, and extrinsic curvature, $K_{ij}$, over time \cite{2007gr.qc.....3035G}. A free evolution scheme is employed, meaning the Hamiltonian and momentum constraints are used solely as diagnostics to assess the accuracy of our numerical method after the initial data has been constructed. We restrict our analysis to spherically symmetric black holes, using the standard polar spherical topology with spatial coordinates $x^i = (r, \theta, \phi)$. 

To update the gauge functions, we choose the generalized Eddington-Finkelstein slicing condition,  $\alpha/\sqrt{\gamma_{rr}} + \beta^r = 1$, where the lapse function $\alpha$ follows the Eddington-Finkelstein requirement that $t + r$ is an ingoing null coordinate. Additionally, we select the shift vector $\beta^r$ such that  $-2\alpha K_{\theta \theta} + \beta^r \partial_r \gamma_{\theta \theta} = 0$, which guarantees that $\partial_t \gamma_{\theta \theta} = 0$, i.e., the areas of constant-$r$ surfaces remain constant during evolution \cite{1993PhRvL..70....9C, 1995PhDT........29M, 1996PhRvD..54.4929M}. 

The matter field is modeled as a scalar field, $\Phi$, whose dynamics are governed by the stress-energy tensor, 
$T_{\alpha \beta}=\partial_{\alpha} \Phi ~\partial_{\beta} \Phi-\frac{1}{2} g_{\alpha \beta}~\partial^{\delta}\Phi~ \partial_{\delta}\Phi$. The evolution of the scalar field is determined by the local conservation law, 
$\nabla_{\beta} T^{\alpha \beta} = 0$, which simplifies to the Klein-Gordon equation: $g^{\alpha \beta}\nabla_{\alpha} \nabla_{\beta} = m^2\Phi$, where $m$ represents the mass of the scalar field. To numerically solve this equation and couple it to the spacetime geometry, we employ the 3+1 formalism. This involves decomposing the Klein-Gordon equation into a system of two first-order equations, facilitating its integration into the framework of general relativity. This approach allows us to study the interplay between the scalar field and the evolving spacetime geometry in a computationally tractable manner. 

To evolve the coupled ADM-Klein-Gordon system of equations in spherical symmetry, we employ the \texttt{GRSFsphe} code. This numerical tool has been extensively validated and used to study a variety of astrophysical scenarios, including black hole dynamics and scalar field interactions \cite{PhysRevD.85.024036, 2014MNRAS.443.2242L, 2013JCAP...12..015L}. 

The evolution requires initial data that are consistent with Einstein's equations. To construct such data, we solve the Hamiltonian and Momentum constraints for the metric components
$\gamma_{rr}$ and $K_{\theta \theta}$. We adopt the following ansatz for the initial conditions:   $\gamma_{\theta \theta}=r^2$, $K_{rr}=-\left[2 M / r^2\right] \times$ $[(1+M / r) / \sqrt{1+2 M / r}]$, where $M$ denotes the mass of the black hole. We assume an initial profile corresponding to a spherical wave modulated by a Gaussian distribution for the scalar field. This choice is motivated by cosmological models of quintessence scalar fields \cite{urena2011black}. Specifically, the scalar field profile is given by:
\begin{equation}
 \Phi(r)=A \frac{\cos (k r)}{r} e^{-\left(r-r_0\right)^2 / \sigma^2},
 \label{eq:scalaerfield}
\end{equation}
where it is centered at  $r=r_0$, with amplitude $A$,  wave number $k$, and width packet $\sigma$. It is important to note that this initial data does not represent an equilibrium solution. The numerical details of the code and the analysis of the black hole scalar field system dynamics can be found in \cite{PhysRevD.85.024036,2013JCAP...12..015L}.

\begin{figure}
\centering
\includegraphics[width=0.45\textwidth]{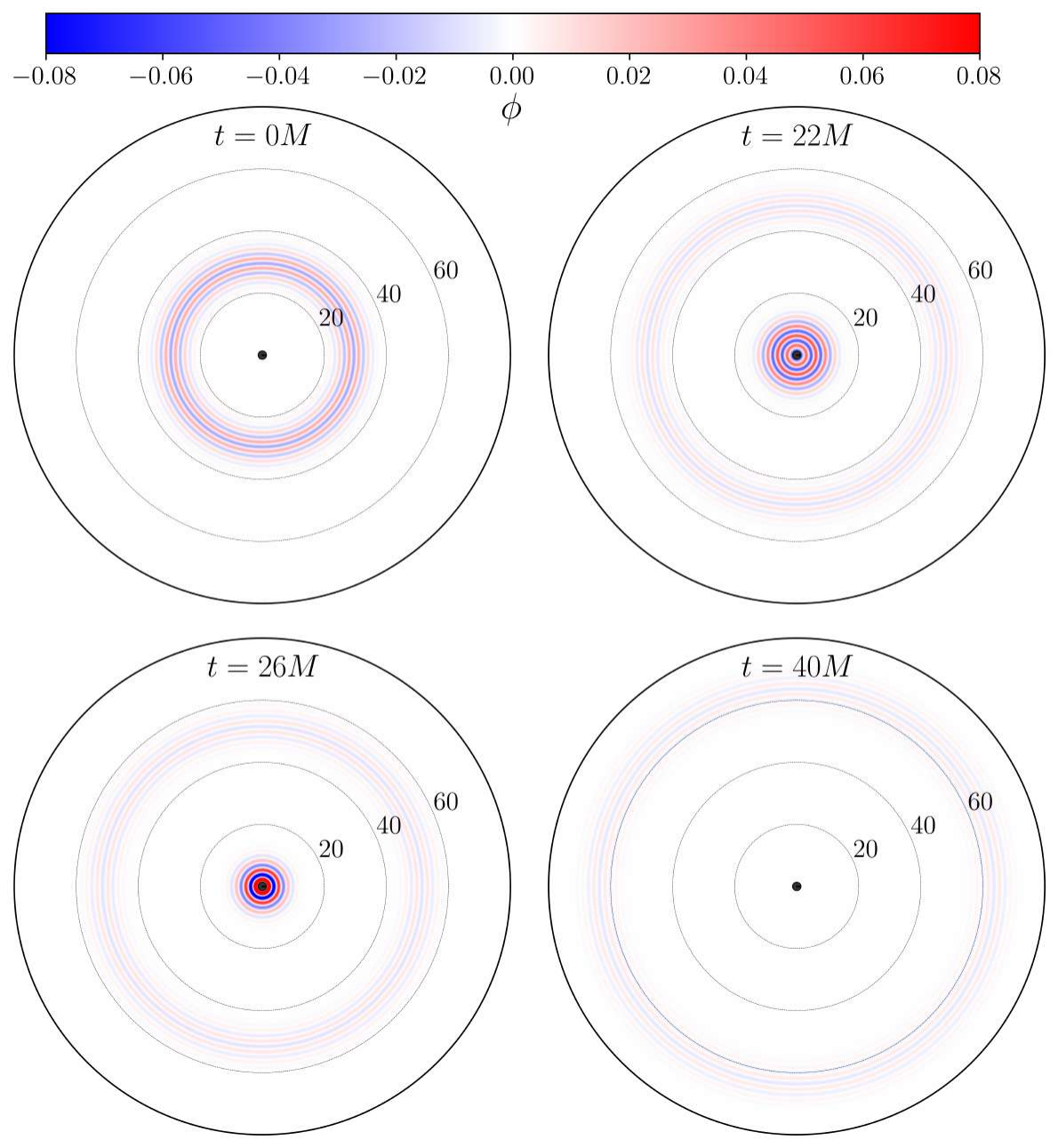}
\caption{\label{fig:scalarfield} 
Time evolution of the scalar field interacting with the black hole. The initial configuration for the scalar field is characterized by $k = 2,\ \sigma = 5,\ A = 0.8$, and  $r_0 = 30\, M$, corresponding to the dark star scenario depicted in Figure \ref{fig:photonring}. The four panels display snapshots of the nonlinear evolution of the scalar field at times $t=0\,M, 22\,M, 26\,M, 40\,M$, as governed by equation \eqref{eq:scalaerfield}. The simulations were performed at the highest resolution, $\Delta r_{3}$, ensuring numerical accuracy.}
\end{figure}

Figure \ref{fig:scalarfield} illustrates the evolution of the initial scalar field pulse, which splits into two distinct components: one propagating inward toward the black hole and the other moving outward. To highlight the dynamics of the black hole-scalar field system, we present a representative case from our simulations with parameters $k = 2$, $\sigma = 5$, $A = 0.8$, and  $r_0 = 30\, M$. Notably, the amplitude of the scalar field rapidly stabilizes around zero in the region left behind by the outward-moving pulse. 

\begin{figure}
\centering
\includegraphics[width=0.45\textwidth]{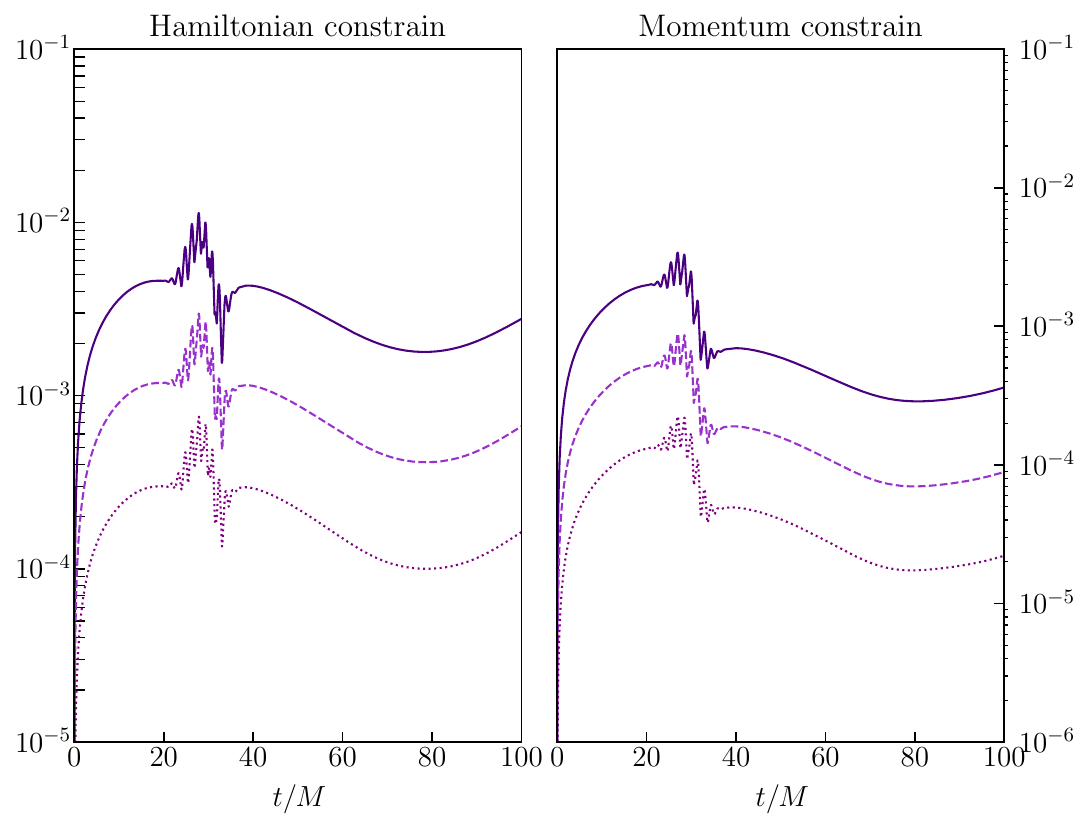}
\caption{\label{fig:Convergencia} $L_2$ norm of the Hamiltonian and Momentum constraints as a function of time. The initial data for the scalar field are described in Figure \ref{fig:scalarfield}. In each panel, we show the amplitude of the constraint violation when we increase the numerical resolution: solid lines for $\Delta r_{1}$, dashed lines for $\Delta r_{2}$, and dotted lines for $\Delta r_{3}$, respectively. 
}
\end{figure}

To validate the accuracy of our numerical results, we demonstrate second-order convergence, confirming that constraint violations diminish as the resolution increases. In Figure \ref{fig:Convergencia}, we show the convergence of the $L_2$ norm of the constraint violations toward zero for the simulation depicted in Figure \ref{fig:scalarfield}. The simulations were performed at three resolutions: $\Delta r_{1} = 0.0375\, M$ (solid line), $\Delta r_{2} = \Delta r_{1}/2$ (dashed line), and $\Delta r_{3} = \Delta r_{2}/2$ (dotted line). The observed convergence behavior underscores the reliability of our numerical approach.


\section{Shadow and gravitational lensing} \label{sec:results}

We carried out numerical simulations using \texttt{OSIRIS} ({\bf O}rbits and {\bf S}hadows {\bf I}n {\bf R}elativ{\bf I}stic {\bf S}pace-times) \cite{2022EPJC...82..103V,2023MNRAS.519.3584V}, a code of our authorship, based on the backward ray-tracing algorithm for stationary and axially symmetric space-times. We generalize \texttt{OSIRIS} to handle dynamic spacetimes, enabling us to trace the evolution of the shadow produced by the coupled Einstein and scalar field system of equations. \texttt{OSIRIS} evolves null geodesics backward in time by solving the equations of motion in the Hamiltonian formalism. The code is based on the image-plane model, an assumption where photons come from the observation screen in the direction of the compact object. This method is commonly used to simulate shadows around black holes \citep{Johannsen_2013}, classifying photon orbits into two categories: those that reach the event horizon and those that escape to infinity. Each pixel on the image plane corresponds to an initial condition for a photon. To classify the orbits and elucidate gravity's effect on null geodesics, we define a celestial sphere as a bright source from which light rays will be emitted. This sphere is concentric with the black hole-scalar field system, surrounds the observer, and is divided into four colors: blue, green, yellow, and red. The notion of curvature is provided by a black mesh with constant latitude and longitude lines, separated by $6^\circ$ (see \cite{2022EPJC...82..103V} for more details of the orbits classification). The shadow produced by this system is viewed by a Minkowskian observer located at the equatorial plane, at a distance of $r = 110\, M$. Finally, the observation screen has a range of $-40 M \le x, y \le 40 M$ with a resolution of $1024\times 1024$ pixels for all simulations. 

\begin{figure}[]
\includegraphics[width=0.45\textwidth]{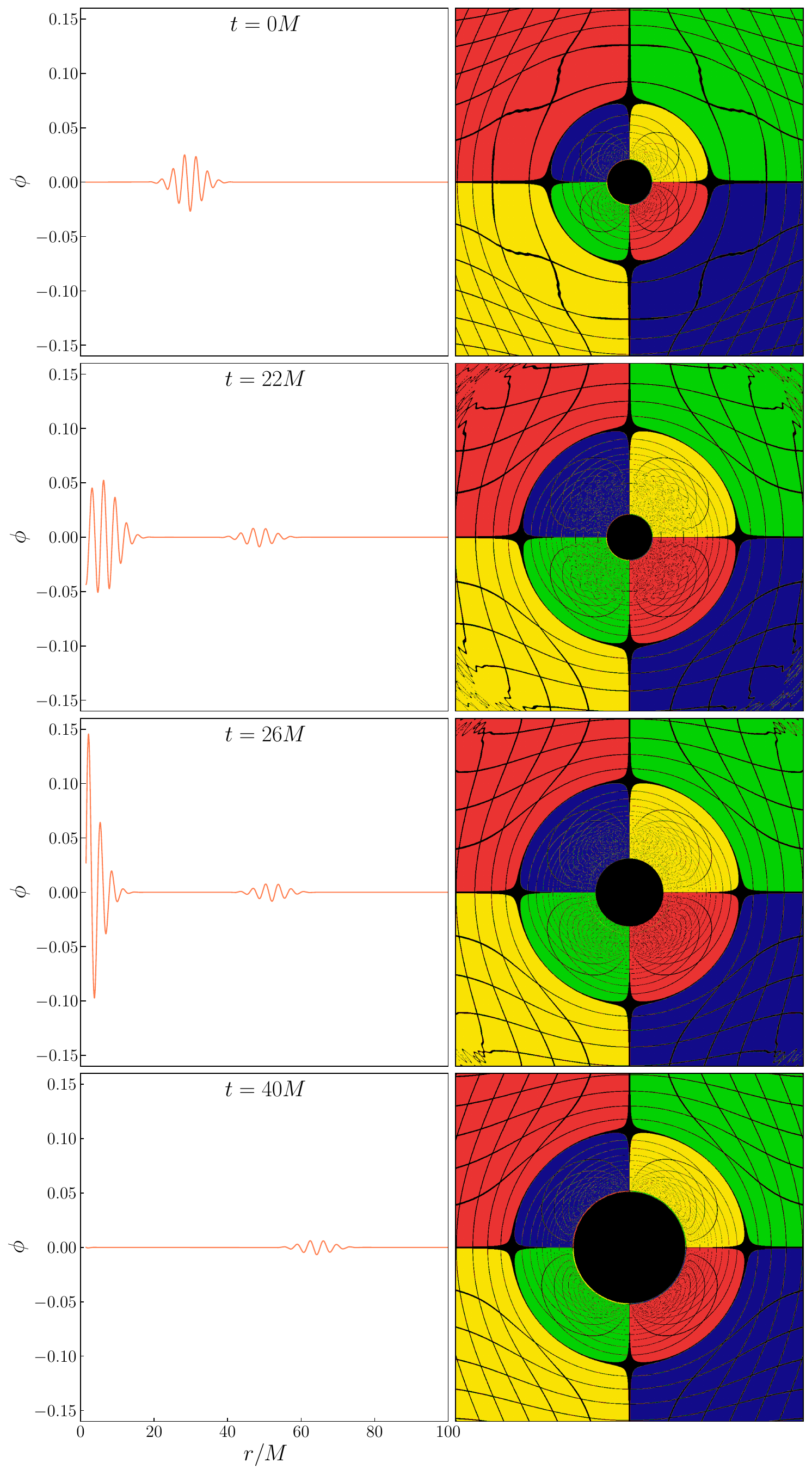}
\caption{\label{fig:evolution} Nonlinear co-evolution of the scalar field, the dynamically growing black hole shadow, and gravitational lensing. The left panels show a sequence of scalar field profiles over time, up to $t=40\,M$ when the system reaches a quasi-steady state. The right panels show the evolution of the black hole shadow, with the profile at $t=0\,M$ being Schwarzschild (same as in Figure \ref{fig:scalarfield}), growing as the black hole mass increases by accreting the scalar field.
}
\end{figure}

In Figure \ref{fig:evolution} we present a series of snapshots of the dynamical evolution of the gravitational lensing produced by the self-gravitating black-hole scalar field system. The set of parameters is the same as in Figure \ref{fig:scalarfield}. The Einstein ring is an appreciable effect of the gravitational lens, a phenomenon that appears when the observer, the compact object, and the source are aligned. As a result, the light source looks like a concentric ring around the black hole. Inside this ring, the deflection angle of the photons' trajectories is large enough to cause an inversion of the colors.  A second ring can be seen near the edge of the shadow, where the images are inverted again. Remarkably, as the scalar field pulse moving towards the black hole is accreted, an increase in the size of both the shadow and the Einstein ring is evident. For this particular set of parameters, the size of these general relativistic effects increases by up to $2.42$ and $1.5$ times their initial values, respectively. It is worth mentioning that the initial size of the shadow is $1.02$ times that of the Schwarzschild solution. This slight difference is due to the gravitational field produced by the scalar field.

\begin{figure}
\centering
\includegraphics[width=0.45\textwidth]{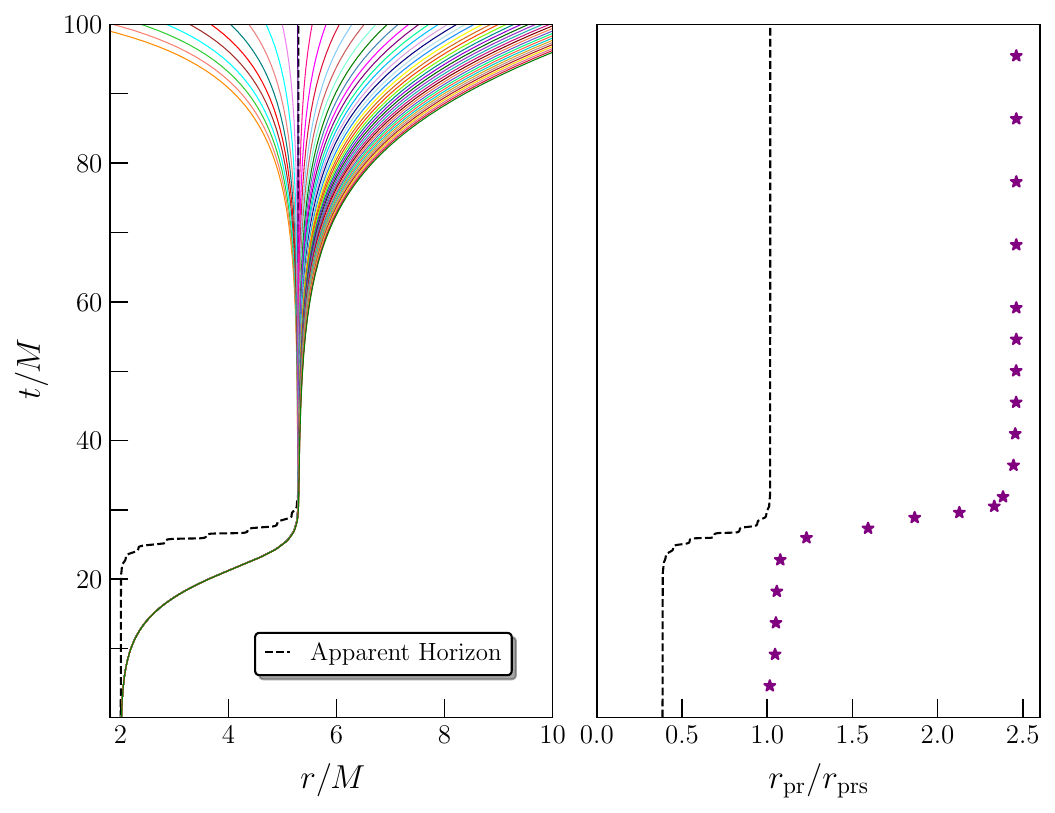}
\caption{\label{fig:shadowsize} 
Evolution of the black hole event horizon, apparent horizon, and photon ring size for the model described in Figure \ref{fig:scalarfield}. Left: The dashed line represents the apparent horizon evolution, while the colored lines show the congruence of outgoing null geodesics, indicating the event horizon location. Right: Comparison between the apparent horizon and photon ring (purple stars) size during the evolution, showing consistent correlation; {\it i.e.}, the photon ring grows within the apparent horizon.
}
\end{figure}

To establish that the expansion of the black hole area is not merely a gauge artifact, the left panel of Figure \ref{fig:shadowsize}, illustrates a fine-tuned bundle of outgoing null rays showing the location of the event horizon. Importantly, we emphasize the gauge invariance of the event horizon. Additionally, our observations consistently confirm that the apparent horizon always lies within the boundaries of the event horizon \cite{hawking1973large}. On the other hand, in the right panel of figure \ref{fig:shadowsize}, we observe the dynamical evolution of the photo-ring radius $r_{\rm pr}$ normalized by Schwarzschild radius $r_{\rm prs}$. The most substantial expansion of the photon ring occurs between times $t=25\, M$ and $t=30\, M$, coinciding with the near-total absorption of the scalar field. After accretion, $r_{\rm pr}$ stabilizes at approximately $r_{\rm pr}=12.8\,M$, which is about $2.46$ times Schwarzschild photon ring $r_{\rm prs}$. Given that the photon ring is an observable phenomenon, our findings are significant for comparative analyses with Sgr\,A*, and M87* observations. 
From an observational standpoint, regardless of whether the underlying solution is a non-rotating spherically symmetric or axisymmetric rotating one, the shadow size of M87\,* falls within the range $4.31\,M \le r_{\rm prs} \le 6.08 \,M$ \cite{EHT_M87_PaperI, EHT_M87_PaperVI}, while for the galactic center Sgr\,A*, the shadow range is $4.5\,M \le r_{\rm prs} \le 5.5\, M$ \cite{EHT_SgrA_PaperI, EHT_SgrA_PaperVI}.

\begin{figure}
\centering 
\includegraphics[width=0.45\textwidth]{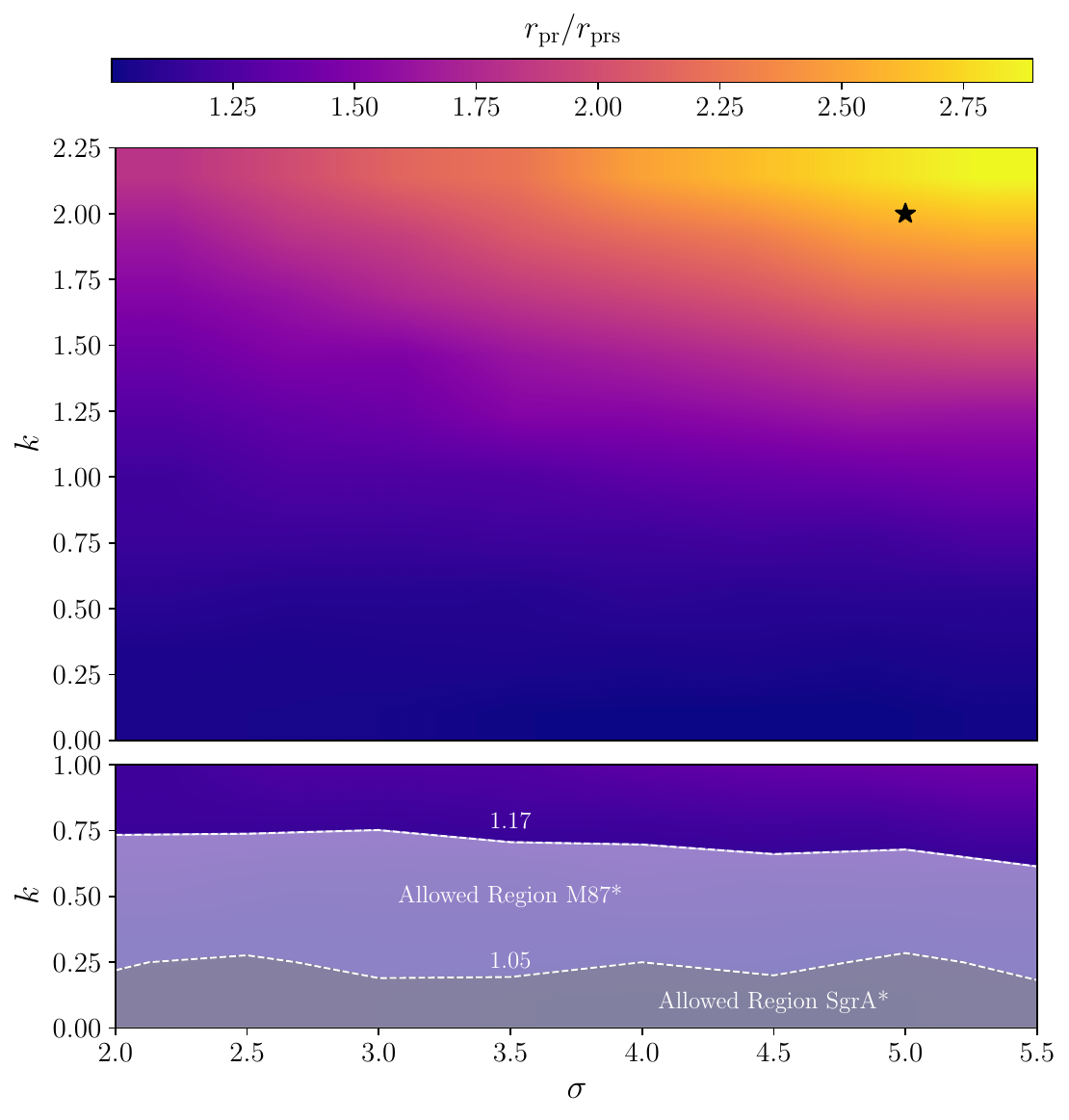}
\caption{\label{fig:photonring} Photon ring size normalized by the Schwarzschild black hole photon ring $r_{\rm prs}$ as a function of wave number $k$ and packet width $\sigma$, measured at the steady state after $t\sim 40\, M$ in all models. We have explored $80$ models. The observational size of the photon ring for the supermassive black hole M\,87 can be reproduced by models in the shaded region for $k\lesssim 0.75$, while the photon ring for SgrA* can be reproduced by models in the region of $k\lesssim 0.25$. The dark star marks a representative case discussed in previous figures.}
\end{figure}

In Figure \ref{fig:photonring}, we present the final photon ring once the accretion process has stopped, normalized by the Schwarzschild photon ring for the different values of the $k$ and $\sigma$ parameters. This Figure reveals a noteworthy trend: the photon ring radius closely resembles that of the Schwarzschild one for $ k < 1.0$, regardless of the chosen $\sigma$. Nonetheless, as both $\sigma$ and $k$ increase, the distinction between radius sizes becomes more pronounced. In particular, when $\sigma = 5.5$ and $k = 2.25$, the most significant expansion occurs when the scalar field is accreted, resulting in a photon ring radius of $15\,M$. This measurement is approximately $2.8$ times greater than the photon ring radius produced by a Schwarzschild black hole. The black star corresponds to the case in Figure \ref{fig:evolution} and \ref{fig:shadowsize}. In this particular case, the black hole’s final mass grows to $2.4,M$ after accreting a scalar field mass of $M_{\phi} = 1.4\,M$, leading to an expansion of the photon ring to 2.4 times the size of the Schwarzschild photon ring. This growth clearly exceeds observational constraints of $3\sqrt{3}(1 \pm 0.17), M$ \cite{Psaltis2020,Kocherlakota2021}. Nevertheless, we explore a broad range of scalar field configurations in this work by varying the wave number and packet width, $k$, and $\sigma$, enabling us to reproduce the observed shadow size.

The bottom panel of Figure \ref{fig:photonring} presents our results compared with observations of the supermassive black holes M87* and Sgr\,A*. Our results agree with the observation of M87* for approximate values of $k \lesssim 0.75$ when the accreted mass of the scalar field mass is $M_{\phi} \lesssim 0.17\,M$, while for Sgr\,A*, this agreement holds for $k \lesssim 0.25$ and the accreted mass of the scalar field $M_{\phi} \lesssim 0.05\,M$. 
The self-gravitating black hole–scalar field system offers a compelling alternative framework for understanding the dynamics of Sgr A$^*$. Rather than requiring a rapidly rotating black hole, the observed properties of Sgr\,A* could be explained by the intrinsic gravitational effects of such a system. This interpretation is consistent with constraints on the spin parameter, which suggest 
$a_{\star} \leq 0.1$, as inferred from the spatial distribution of the S-stars \cite{2020ApJ...901L..32F,2022ApJ...932L..17F}.

Finally, an analytical expression for the photon-ring size in terms of the scalar field wave packet parameters $k$ and $\sigma$ is obtained after fitting our numerical results
\begin{equation}
    \log\left(\frac{r_{\rm pr}}{r_{\rm prs}}\right) = \left( a + b k + c k^{2}\right)\left(d + e\sigma \right),
\end{equation}
where the values of the fitting coefficients are: 
$a = -9.65\times 10^{-4}$, $a = 1.54\times 10^{-2}$, $c = 6.11\times 10^{-3}$, $d = 5$ and $e = 2.32$. This expression shows that the photon ring radius increases exponentially with the wave number and packet width. Using this expression, it is possible to obtain the photon ring ratio $r_{\rm pr}/r_{\rm prs}$ when considering different initial values of $k$ and $\sigma$ as long as they are not very different from those considered here.

\section{\label{sec:discussion} Discussions and Conclusions}

In this work, we present the evolution of the shadow and gravitational lensing effects resulting from the self-consistent accretion of a scalar field onto a non-rotating black hole by solving simultaneously the Klein-Gordon-Einstein field equations and the ray-tracing equations.

As the scalar field pulse moves toward the black hole and is gradually accreted, the black hole mass and the apparent horizon increase. This interaction leads to a noticeable expansion in both the photon ring and the Einstein ring; this expansion reflects the dynamical response of the black hole’s gravitational lensing, providing insights into the complex interplay between black holes and scalar fields. The results presented here show that by selecting an appropriate combination of scalar field initial parameters -- specifically, the wave number $k$ and the packet width $\sigma$ -- we can reproduce the shadow sizes of Sgr\,A* and M87 as observed by the Event Horizon Telescope collaboration, constraining the wave number to values $k\lesssim 0.75$ and $k\lesssim 0.25$, respectively.

Our results suggest that the observed shadow may provide valuable insights into the gravitational interactions between the black hole and the surrounding scalar field dark matter. However, our findings could be further refined in future work by incorporating the proposed effects of black hole spin, alternative models of the scalar field, and the impact of the self-gravitating absorption of photons, as has been recently proposed in \citep{DiFilippo2024}. 

In astrophysics, dynamical timescales are primarily determined by the mass of the black hole. For the supermassive black holes considered in this work, Sgr A$^*$ has a mass of $4.14 \times 10^6 \, M_{\odot}$, while M87$^*$ has a mass of $6.5 \times 10^9 \, M_{\odot}$, respectively. The dynamical timescale of the scalar field accretion in geometrized units of $40\, M$, corresponds to $\sim 14$ minutes for Sgr\,A$^*$, and $\sim 14$ days for M87\,$^*$. Recent observational campaigns by the Event Horizon Telescope (EHT) collaboration, spanning six days \cite{EHT_M87_PaperI, EHT_SgrA_PaperI}, and by the Atacama Large Millimeter Array (ALMA), covering 15 hours \cite{2022ApJ...930L..19W}, enable the study of Sgr\,A$^*$ light curves at 230GHz on timescales ranging from minutes to hours \cite{EHT_SgrA_PaperIV}. Additionally, for M87$^*$, the EHT collaboration has conducted multiyear observations \cite{EHT_M87_2018_2024}, with each campaign covering periods between two and six days. Assuming that the scalar field cloud is accreting during the observation campaigns, our analysis suggests that the dynamical effects of the scalar field on black hole shadows could be detectable. Although identifying such events is challenging, their observation remains a plausible possibility within the capabilities of current instruments. Moreover, future observations with improved sensitivity and resolution will enhance the prospects of detecting these effects.


\begin{acknowledgments}
We thank Luciano Rezzolla for his valuable discussions and comments on the manuscript.
F.D.L-C is supported by the Vicerrectoría de Investigación y Extensión - Universidad Industrial de Santander, under Grant No. 3703. J.C.A-M deeply thanks Universidad Industrial de Santander for the forgivable loan support.
A.C.O gratefully acknowledges to PAPIIT DGAPA-UNAM project IA103725, and ``Ciencia Básica y de Frontera 2023-2024" program of CONAHCYT México, projects CBF2023-2024-1102 and 257435.
\end{acknowledgments}

\bibliography{apssamp}

\end{document}